\documentclass[twocolumn,superscriptaddress,aps,prb,showpacs,showkeys,amsmath,preprintnumbers]{revtex4}

\usepackage{graphicx}												
\usepackage{siunitx}	
\usepackage{scrtime}												

\makeatletter
\DeclareRobustCommand*\textsubscript[1]{%
  \@textsubscript{\selectfont#1}}
\newcommand{\@textsubscript}[1]{%
  {\m@th\ensuremath{_{\mbox{\fontsize\sf@size\z@#1}}}}}
\makeatother  

\newcommand{\Molek}[2]{{#1\textsubscript{{#2}}}}										
\newcommand{\NbFe}{\Molek{NbFe}2}
\newcommand{\NbFey}{\Molek{Nb}{1-y}\Molek{Fe}{2+y}}
\newcommand{\NbFeFe}{\Molek{Nb}{0.985}\Molek{Fe}{2.015}}
\newcommand{\NbFeNb}{\Molek{Nb}{1.01}\Molek{Fe}{1.99}}

\newcommand{\RH}{\ensuremath{R_{\mathrm H}}}								
\newcommand{\RHn}{\ensuremath{R_{\mathrm H}^{\mathrm n}}}								
\newcommand{\SH}{\ensuremath{S_{\mathrm H}}}								
\newcommand{\RS}{\ensuremath{R_{\mathrm S}}}								
\newcommand{\RT}{\ensuremath{\RH (T)}}								
\newcommand{\rhoH}{\ensuremath{\rho_{\mathrm H}}}				
\newcommand{\TC}{\ensuremath{T_{\mathrm C}}}
\newcommand{\TN}{\ensuremath{T_{\mathrm N}}}

\newcommand{\HEunit}[1]{\SI{#1 e-10}{\meter\cubed\per\coulomb}}

\begin{document}

\preprint{Manuscript: version 0.2 Date: \today\  \thistime}

\title{Normal and intrinsic anomalous Hall effect in \NbFey}

\author{Sven Friedemann}
	\email{sf425@cam.ac.uk}
\affiliation{Cavendish Laboratory, University of Cambridge, JJ Thomson Avenue, CB3 0HE Cambridge, UK}

\author{Manuel Brando}

\affiliation{Max Planck Institute for Chemical Physics of Solids, N\"othnitzer Strasse 40, 01187 Dresden, Germany}

\author{William J Duncan}

\affiliation{Department of Physics, Royal Holloway, University of London, Egham TW20 0EX, United Kingdom}

\author{Andreas Neubauer}
\author{Christian Pfleiderer}

\affiliation{Physik Department E21, Technische Universit\"at M\"unchen, James-Franck-Strasse, D-85748 Garching, Germany}

\author{F Malte Grosche}

\affiliation{Cavendish Laboratory, University of Cambridge, JJ Thomson Avenue, CB3 0HE Cambridge, UK}

\date{\today }

\begin{abstract}
The Hall effect on selected samples of the dilution series \NbFey\ is studied. Normal and anomalous contributions are observed, with positive normal Hall effect dominating at high temperatures. Consistent analysis of the anomalous contribution is only possible for Fe-rich \NbFeFe\ featuring a ferromagnetic ground state. Here, a positive normal Hall coefficient is found at all temperatures with a moderate maximum at the spin-density-wave transition.  The anomalous Hall effect is consistent with an intrinsic (Berry-phase) contribution which is constant below the ordering temperature \TC\ and continuously vanishes above \TC. For stoichiometric \NbFe\ and Nb-rich \NbFeNb\ -- both having a spin-density-wave ground state -- an additional contribution to the Hall resistivity impedes a complete analysis and indicates the need for more sophisticated models of the anomalous Hall effect in itinerant antiferromagnets.
\end{abstract}
\maketitle

%
%
\section{Introduction}
The Hall effect marks the generation of a transverse voltage $V_y$ arising from the deflection of moving charge carriers due to the Lorentz force in a magnetic field $H$ perpendicular to the driving current $I$ \cite{Hurd1972}. 
The Hall coefficient defined as $\RH = V_y/(t I \mu_0 H)$ is a powerful probe of the Fermi surface of metals, and as such is particular interesting for the study of electronic instabilities in metals \cite{Friedemann2011}. Moreover, measurements of the Hall resistivity \rhoH\ provide access to scattering rates and charge carrier mobilities. Here, $t$ denotes the thickness of the sample.

Shortly after the initial discovery of the Hall effect, E.H.~Hall found the anomalous Hall effect (AHE) in magnetic materials \cite{Hall1879a}. The AHE stems from spin-orbit coupling \cite{Nagaosa2010}. It is both of fundamental interest and has the prospect for important applications as it allows to manipulate charge and spin degrees of freedom \cite{Wolf2001,Toyosaki2004}. The anomalous contribution \RS\ adds to the normal Hall coefficient \RHn\ in the Hall resistivity
\begin{equation}
	\rhoH = \RHn \mu_0 H +  \RS M
\label{eq:AHEI}
\end{equation}
This empirical description of the anomalous Hall effect is based on the spontaneous magnetisation and is established for a broad class of materials. 

Two classes of theoretical approaches model the AHE microscopically: On the one hand, the intrinsic AHE is concisely seen using topological concepts, the Berry phase \cite{Berry1984}, and is an important extension of the Fermi-liquid concept \cite{Haldane2004}. The extrinsic AHE, on the other hand, arises from asymmetric (skew) scattering processes \cite{Smit1955,Smit1958}. The skew scattering picture predicts the anomalous contribution to scale linearly with the resistivity $\RS \propto \rho$.

The intrinsic AHE is related to other topological concepts like the topologically protected states in quantum Hall systems and may itself be quantized in some cases \cite{Yu2010}. The microscopic theory by Karplus and Luttinger relates the AHE to the band structure and is  predicted to be proportional to the square of the resistivity, $\RS \propto \rho^2$ with the scale factor \SH\ defined via
\begin{equation}
\rhoH (H) = \RHn \mu_0 H  + \SH  \rho^2  M 
\label{eq:AHEII}
\end{equation}

Consequently, intrinsic and skew scattering contributions can be distinguished by their relation to the resistivity. The second scattering process, side jump scattering, can be viewed as a consequence of the intrinsic AHE and as such leads to the same quadratic dependence on resistivity. This explains the independence of the intrinsic contribution to moderate impurity concentrations in the metallic regime with $\rho \sim \SI{e-4} - \SI{e-6}{\ohm\centi\meter}$ \cite{Miyasato2007}.

\begin{figure}%
\includegraphics[width=.7\columnwidth]{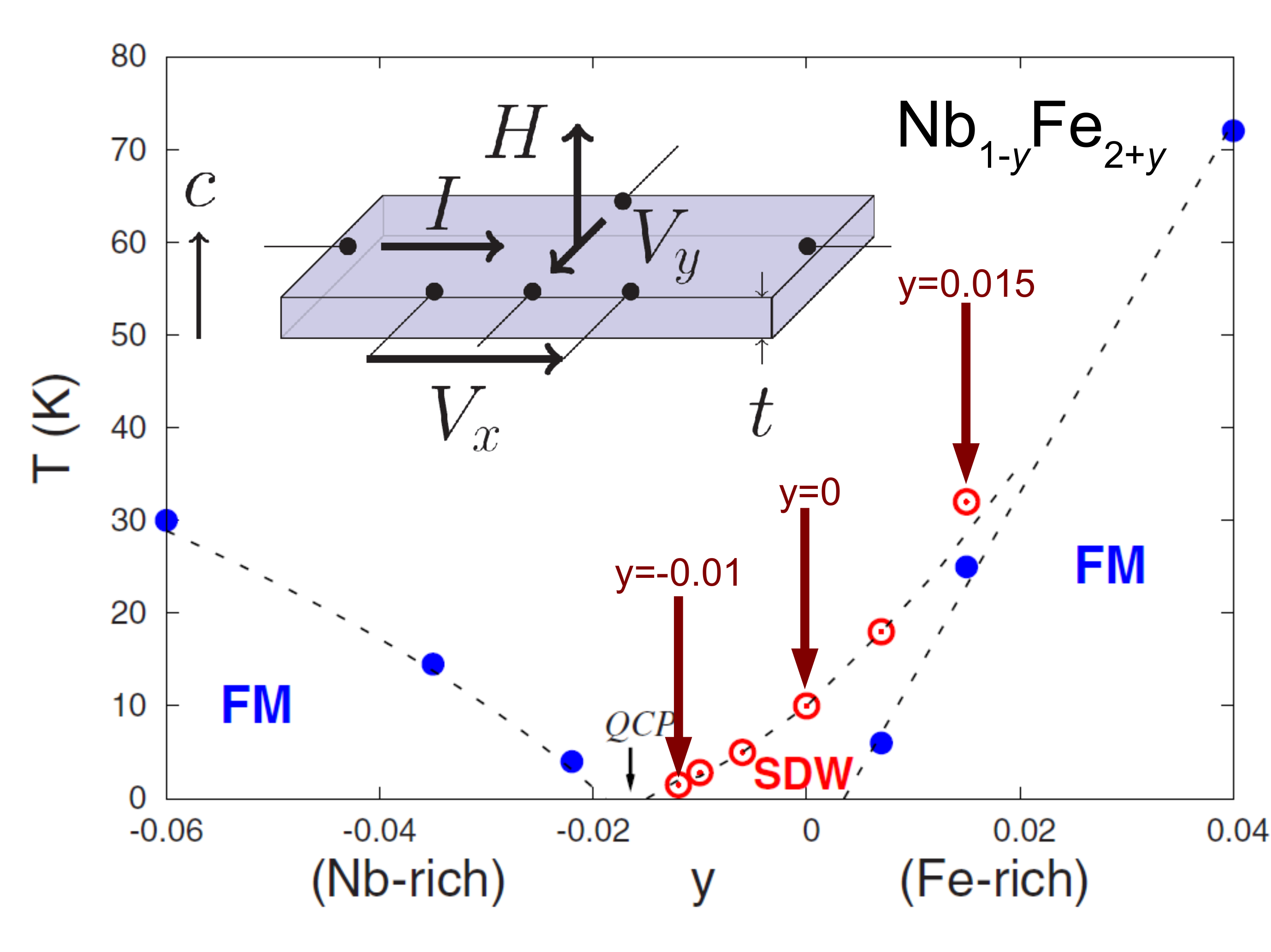}%
\caption{Phase diagram of \NbFey\ after  Ref.~\onlinecite{Moroni-Klementowicz2009}. Ferromagnetism occurs below \TC\ (solid blue circles) for both Fe-rich and strongly Nb-rich samples. On approaching stoichiometry ($y=0$) \TC\ is continuously suppressed to zero temperature. A new, presumably SDW type, phase emerges at temperatures above \TC\ for samples with $ y\leq0.02$. The SDW phase transition at \TN\ (open red circles) is itself suppressed on decreasing the iron content and yields to a QCP at $y \sim -0.01$. Arrows indicate the three samples selected for Hall effect measurements. The inset illustrates sample orientation for Hall effect and resistivity measurements.}%
\label{fig:PD}
\end{figure}

The intermetallic compound \NbFe\ falls into this metallic regime: Single crystals feature residual resistivities in the \SI{10}{\micro\ohm\centi\meter} range. Moreover, as the composition is fine-tuned in the dilution series \NbFey\ with $\SI{-6}{\percent} \lesssim y \lesssim \SI{4}{\percent}$ the magnetic ground state can be modified, as summarized in Fig.~\ref{fig:PD}. Both Fe-rich and strongly Nb-rich samples feature ferromagnetic order at lowest temperatures whereas ferromagnetism is suppressed close to stoichiometry. Approaching from the Fe-rich side a new phase, presumably of spin-density-wave (SDW) type emerges on top of the ferromagnetic phase. The suppression of this SDW-phase induces a quantum critical point (QCP) -- a phase transition at zero temperature -- in slightly Nb-rich \NbFe. This QCP appears to be the origin of a logarithmic Fermi liquid breakdown \cite{Brando2008}. With this broad range of ground states accessible, \NbFey\ offers the prospect to study the relation of the AHE with the type of magnetism and magnetic fluctuations. 

Here, we present Hall effect measurements on selected samples of the dilution series \NbFey\ with $y= -0.01$, $y=0$, and $y=0.015$ (cf.\ phase diagram in Fig.~\ref{fig:PD}). These samples show a magnetic transition (SDW) at $\TN = \SI{3}{\kelvin}$, \SI{10}{\kelvin}, and \SI{31}{\kelvin}, respectively. In addition, the iron rich sample ($y=0.015$) orders ferromagnetically below $\TC = \SI{23}{\kelvin}$.
 
%
%
\section{Experimental setup}
Samples were grown with the optical floating zone method from polycrystalline seed rods. The actual stoichiometry was determined from the crystallographic lattice parameters as detailed in Ref.~\onlinecite{Moroni-Klementowicz2009}.
Oriented samples were cut to thin platelets with thickness $t \approx \SI{100}{\micro\meter}$.

The Hall effect and the resistivity of \NbFe\ have been measured with the driving current $I$, the longitudinal voltage $V_x$ and the transversal voltage $V_y$ in the crystallographic hexagonal plane while the magnetic field $H$ was applied along the crystallographic $c$ direction (cf.\ inset of  Fig.~\ref{fig:PD}). The (longitudinal) resistivity was determined from the longitudinal voltage as $\rho = V_x/I$.
The Hall resistivity was calculated as the antisymmetric part of the transversal voltage as
\begin{equation}
\rhoH(H)=t\left[V_y(+H)-V_y(-H)\right]/2I
\label{eq:rhoH}
\end{equation}
This procedure allows to separate components of the magnetoresistance arising from small misalignment of the transversal contacts.
We note that the accuracy of \SI{10}{\percent} in the determination of the thickness applies to all deduced quantities \rhoH, \RHn, and \SH\ and may lead to discrepancies when comparing data taken on different samples.  Moreover, the Hall coefficient was determined at a fixed field as 
\begin{equation}
\RH (H) = \rhoH(H) / \mu_0 H. 
\label{eq:RH}
\end{equation}


Magnetisation measurements were carried out on a larger piece from which the sample for the Hall measurements was cut. The sample was equally oriented with magnetic field along the crystallographic $c$ axis in order to enable the analysis of the anomalous Hall effect. All measurements were conducted in a Quantum Design Physical Property Measurements System.

%
%
\section{Results and discussion}

%
%
\subsection{Temperature dependent Hall coefficient}
%

\begin{figure}%
\includegraphics[width=.9\columnwidth]{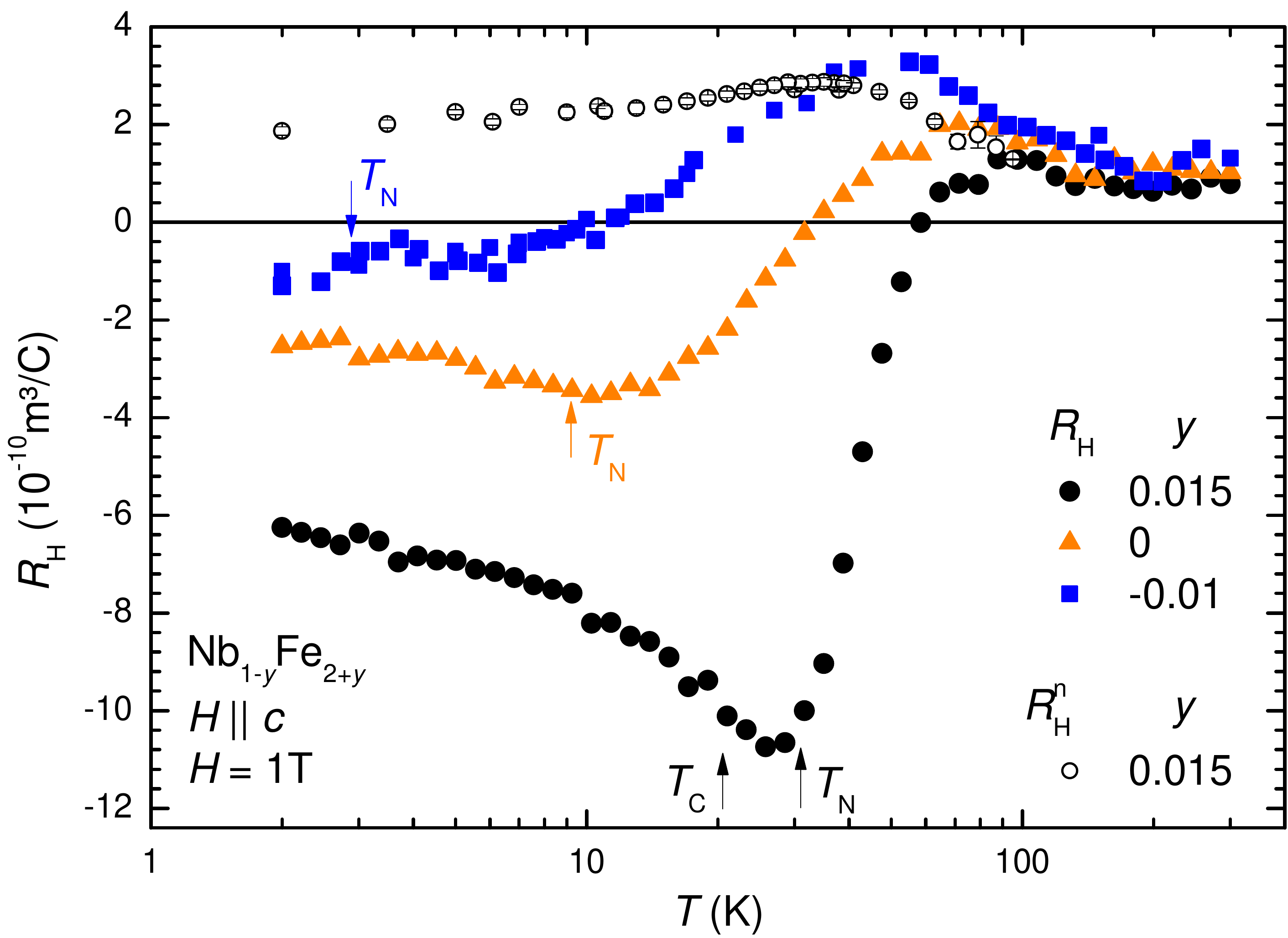}%
\caption{Temperature dependence of the Hall coefficient for samples of \NbFey\ with $y=-0.01$, 0, and 0.015 at a fixed field of \SI{1}{\tesla} as defined in Eq.~\ref{eq:RH}. Open symbols denote the normal Hall coefficient \RHn\ as extracted from fits of Eq.~\ref{eq:AHEII} to $\rhoH(H)$ as shown in Fig.~\ref{fig:RhoH} (a). Arrows mark the (zero-field) transition temperatures \TC\ and \TN.}%
\label{fig:RHvsT_all}
\end{figure}

The temperature dependence of the Hall coefficient at $H=\SI{1}{\tesla}$ is depicted in Fig.~\ref{fig:RHvsT_all} for three samples of  the \NbFey\ dilution series. At high temperatures all samples exhibit a positive Hall coefficient $\RH \approx \HEunit{1}$. As the temperature decreases an increase in \RH\ is evident for all samples, leading to a maximum which is more pronounced for samples with less Fe content: \NbFeNb\ reaches a value of \HEunit{3} while  stoichiometric \NbFe\ and Fe-rich \NbFeFe\ reach values of \HEunit{2} and \HEunit{1}, respectively. Also the maximum appears to be shifted to lower temperature for decreasing Fe content. 
Upon further decrease of the temperature, \RH\ decreases for all samples and eventually becomes negative. For Fe-rich \NbFeFe\ the Hall coefficient drops rapidly and assumes a pronounced minimum around \SI{25}{\kelvin} at absolute values of $\HEunit{-13}$. At temperatures below this minimum, \RH\ increases and tends towards saturation at lowest $T$. For stoichiometric \NbFe\ a minimum in $\RH(T)$ is present slightly above \SI{10}{\kelvin} with the absolute value reduced compared to Fe-rich \NbFeFe. For Nb-rich \NbFeNb\ a minimum is not resolved at temperatures above \SI{2}{\kelvin} and the absolute value of \RH\ is further reduced compared to stoichiometric \NbFe. Consequently, the minimum in \RT\ appears to be linked to the magnetic transition temperature with the absolute value of \RH\ decreasing as the Fe-content decreases.

%
%

\begin{figure*}
\includegraphics[width=.31\textwidth]{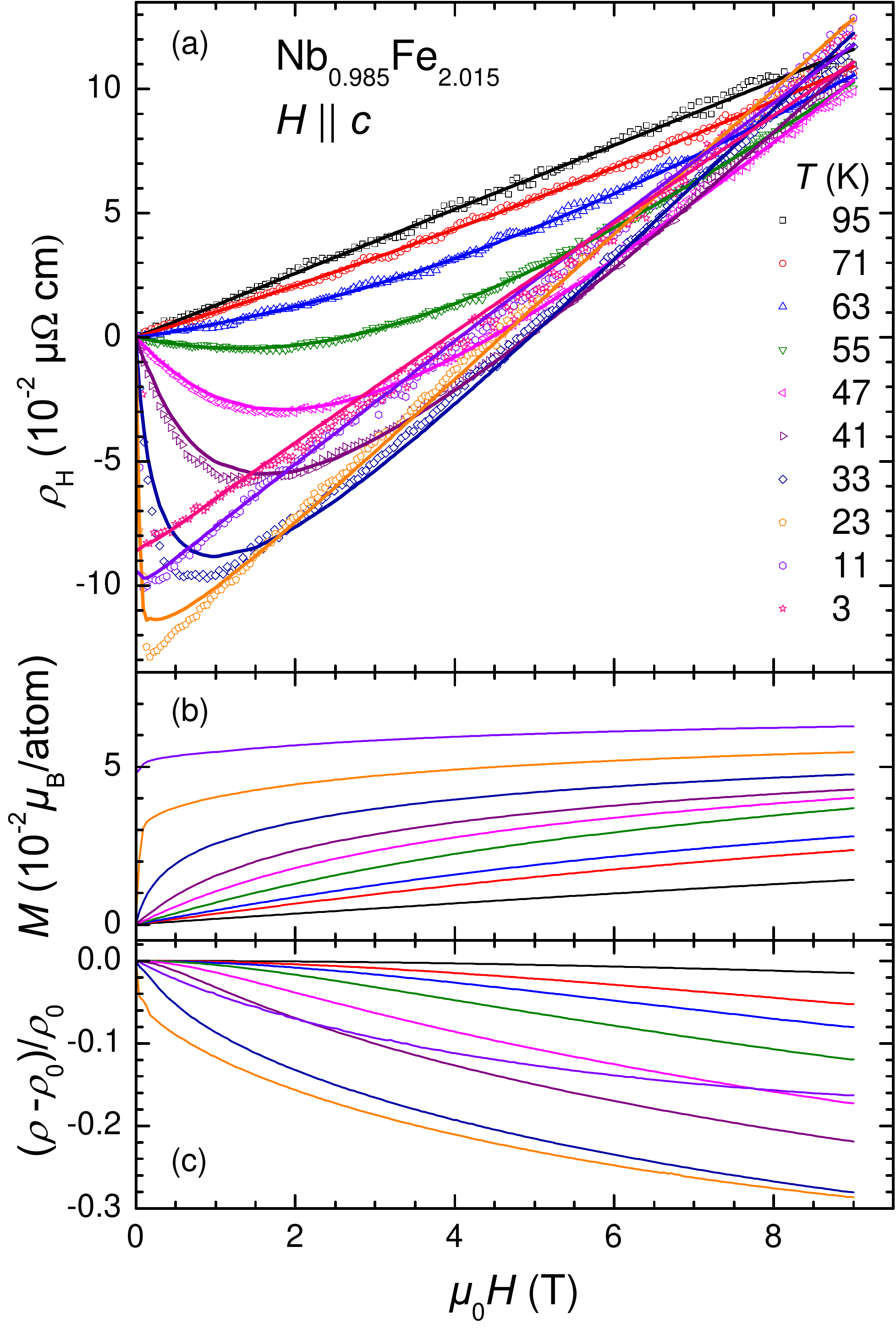}%
\hfill
\includegraphics[width=.31\textwidth]{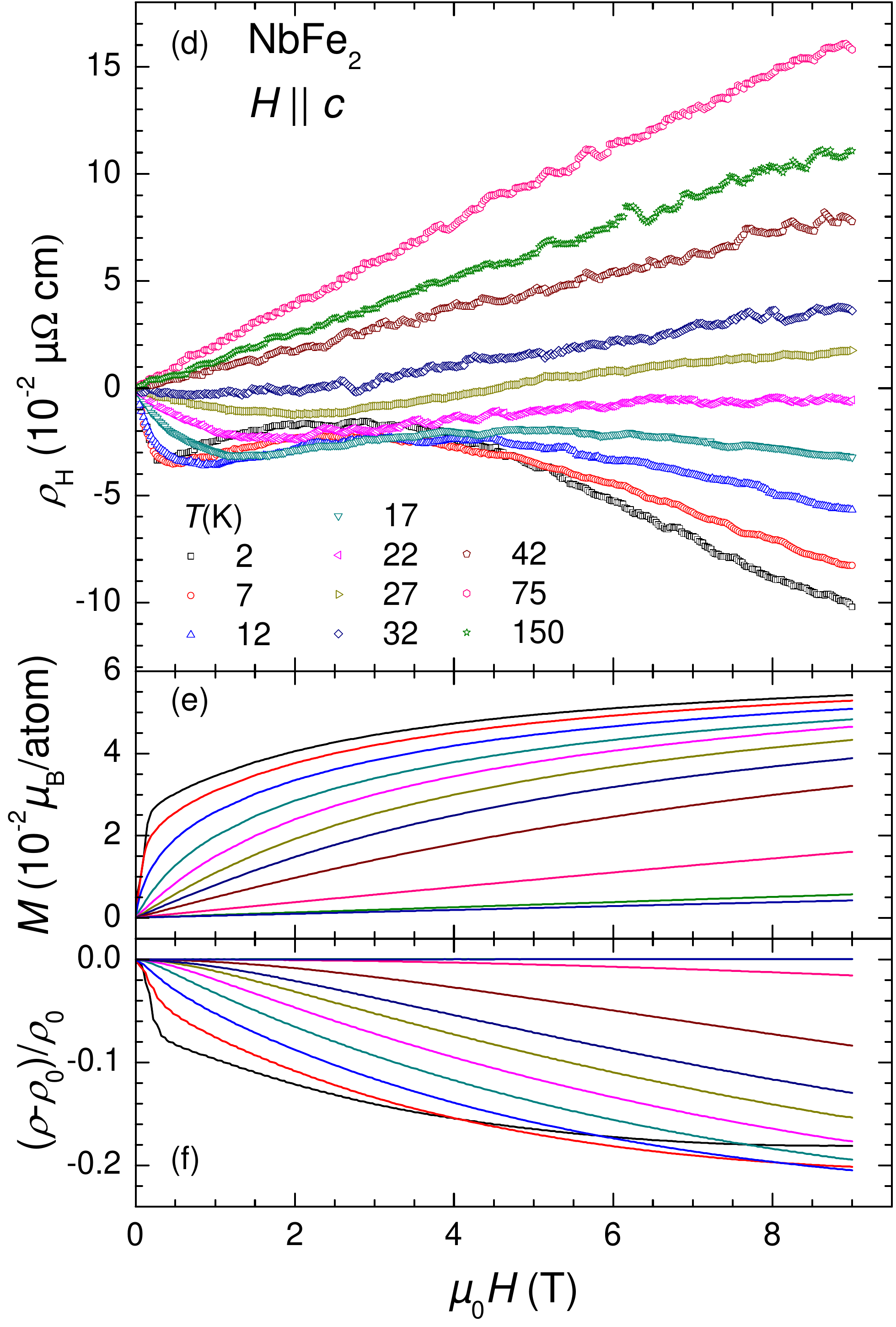}
\hfill
\includegraphics[width=.31\textwidth]{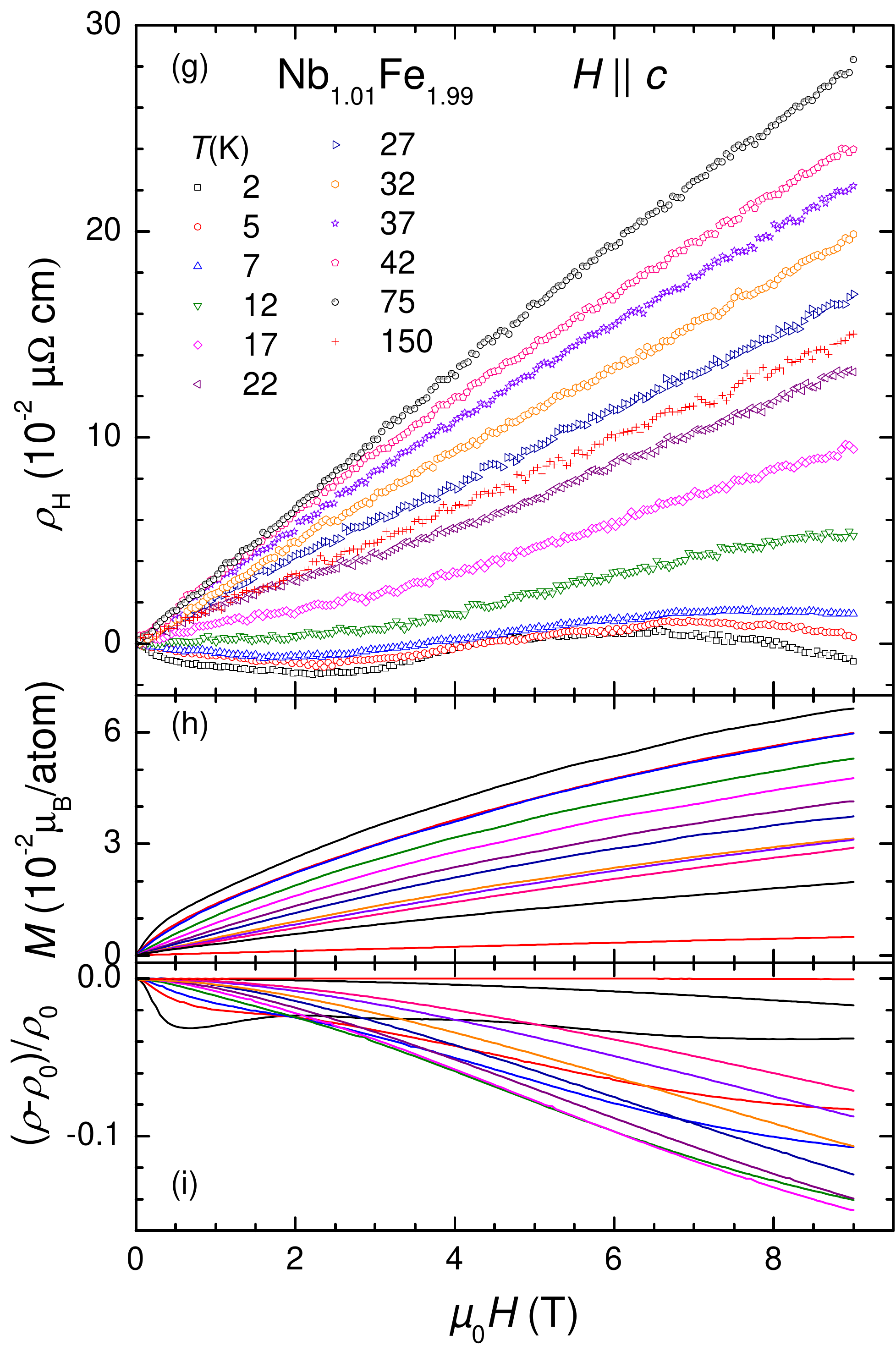}
\caption{Hall resistivity of \NbFey\ as a function of magnetic field for (a) $y=0.015$, (d) $y=0$, and (g) $y=-0.01$. (b), (e), (h)  and (c), (f), (i) depict the Magnetisation and magnetoresistance, respectively. Solid lines in (a) reflect least square fits of $\rhoH$ according to Eq.~\ref{eq:AHEII} utilizing the magnetisation and magnetoresistance data shown in the lower panels.}%
\label{fig:RhoH}
\end{figure*}

\subsection{Field dependent Hall resistivity}
\begin{figure}%
\includegraphics[width=.9\columnwidth]{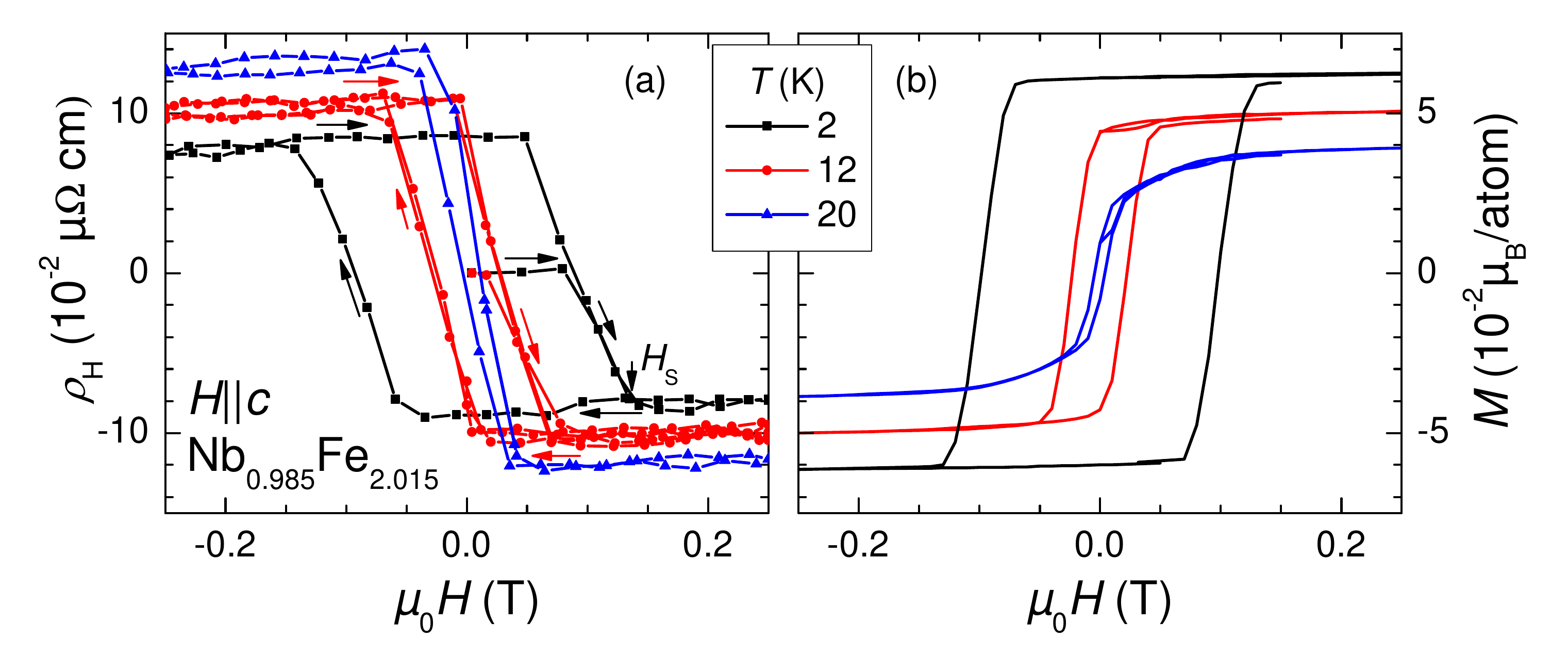}%
\caption{Hysteresis of Hall resistivity (a) and magnetisation (b) in the ferromagnetic phase of Fe-rich \NbFeFe. In order to obtain the Hall resistivity at both positive and negative magnetic field, the Hall resistivity was calculated by subtracting the magnetoresistance contribution such that the hysteresis loops are centred around the origin. }%
\label{fig:OFZ28_Hyst}
\end{figure}

Fig.~\ref{fig:RhoH} (a) displays the field dependence of the Hall resistivity of \textbf{Fe-rich \NbFeFe\ } for selected temperatures. Above \SI{80}{\kelvin} the Hall resistivity $\rhoH(H)$ shows no deviations from linearity. Consequently, here, \RH\ as defined in Eq.~\ref{eq:RH} and depicted in Fig.~\ref{fig:RHvsT_all} is equivalent to the initial-slope Hall coefficient. Below \SI{80}{\kelvin}, however, a concave shape is present in $\rhoH(H)$ which develops to a minimum at temperatures below \SI{55}{\kelvin}. This minimum becomes sharper as temperature is decreased and for temperatures below $\TC=\SI{23}{\kelvin}$ $\rhoH(H)$ resembles two distinct linear sections with a crossover well below \SI{1}{\tesla}, where the initial slope is negative and large while the high-field slope is positive and smaller. This behaviour is typical for the anomalous Hall effect in ferromagnets as shall be discussed below. The two regimes reflect the magnetisation $M(H)$ which shows saturation at about the same field where the slope of the Hall resistivity changes. In fact, the hysteresis of $M(H)$ is also present in the Hall resistivity as can be seen from the comparison of $\rhoH(H)$ and $M(H)$ in Fig.~\ref{fig:OFZ28_Hyst} for $T < \TC$. The hysteresis loops in both quantities obey the same saturation field. The saturation values, however, follow opposed temperature dependence: While the saturation magnetisation increases with decreasing temperature, the saturation value of the Hall resistivity decreases below \TC.
 This decrease in the saturation value of \rhoH\ represents a decrease of the anomalous contribution as expected for a metal with a negative temperature coefficient of the resistivity. 
 
 In order to model the field- and temperature dependence of the Hall resistivity we have tested fits on the basis of skew-scattering and intrinsic AHE. Therefore we have tried fitting forms of eq.~\ref{eq:AHEI} where $R_{\text S}$ was allowed to have different dependencies on resistivity. This detailed analysis reveals best agreement with the data for a relation to $\rho^2$ in agreement with Eq.~\ref{eq:AHEII}. In particular, a linear form of the resistivity was found to produce less poor agreement.

The Hall resistivity of \textbf{stoichiometric \NbFe\ } as depicted in Fig.~\ref{fig:RhoH} (d) has proportionality to the magnetic field $\rhoH \propto H$ for temperatures above \SI{40}{\kelvin} with the slope reflecting the initial-slope Hall coefficient. A concave shape similar to that observed for Fe-rich \NbFeFe\ is present for intermediate temperatures $\SI{40}{\kelvin} \geq T \geq \SI{20}{\kelvin}$. For lower temperatures, however, the Hall resistivity is more complex with a minimum at low fields and a maximum at higher fields. Both these features are shifted to lower fields as the temperature is decreased. Hysteresis is not observed in the Hall resistivity of stoichiometric \NbFe\ above \SI{2}{\kelvin}.

The field dependence of the Hall resistivity of 
\textbf{Nb-rich \NbFeNb\ } is similar to that of stoichiometric \NbFe. As can be seen from Fig.~\ref{fig:RhoH} (g), the Hall resistivity is best described by linear field dependence at high temperatures, however, this is true above \SI{100}{\kelvin} only. In the regime $\SI{100}{\kelvin} \geq T \geq \SI{20}{\kelvin}$, again, the Hall resistivity obeys a concave shape. At lowest temperatures, like for stoichiometric \NbFe, a complex field dependence is observed with a minimum and maximum that shift to lower fields as the temperature is lowered.

%


%
%
\subsection{Analysis}
In order to extract normal and anomalous contributions, the Hall resistivity was fitted utilizing Eq.~\ref{eq:AHEII}. As we shall see, only for Fe-rich \NbFeFe\ this allows a consistent description of the data. The fits for \NbFeFe\ are included in Fig.~\ref{fig:RhoH} (a) as solid lines. The magnetisation and resistivity are shown in panels (b) and (c) respectively. We note, that the magnetisation alone cannot reproduce the shape of the Hall resistivity as the magnetisation shows more pronounced curvature at elevated fields. In other words, Eq.~\ref{eq:AHEI} with a constant \RS\ is not able to fit $\rhoH$. Only a product of the resistivity and the magnetisation leads to a form reproducing the curvature of $\rhoH(H)$ confirming the usage of Eq.~\ref{eq:AHEII}. Remarkable agreement of the fits with $\rhoH(H)$ is  achieved at temperatures above \TN\ and well below \TC\ while small deviations are observed at intermediate temperatures at small fields in the vicinity of the minimum in $\rhoH(H)$.
Here, technical difficulties might impede a better agreement. Most likely small discrepancies of sample orientation during the Hall effect and magnetisation measurements with respect to field can cause differences, but also the fact that $M$ and $\rhoH$ were measured on different pieces of the same sample might play a role \cite{Moroni-Klementowicz2009}. Finally, small differences in sample temperature during magnetisation and transport measurements might contribute to the observed discrepancies. This latter reasoning is supported by the fact that the fits at lowest temperature (see \SI{11}{\kelvin} curve in Fig.~\ref{fig:RhoH} (a)) are again very good, well below \TC\ the magnetisation profile displays less temperature dependence. At lowest temperatures the accurate determination of \SH\ is complicated by the presence of strong hysteresis in both $\rhoH(H)$ and $M(H)$. Here, \SH\ was calculated from the remanent values of \rhoH, $M$, and $\rho$ at zero field deduced from the hysteresis loops.

\begin{figure}%
\includegraphics[width=.9\columnwidth]{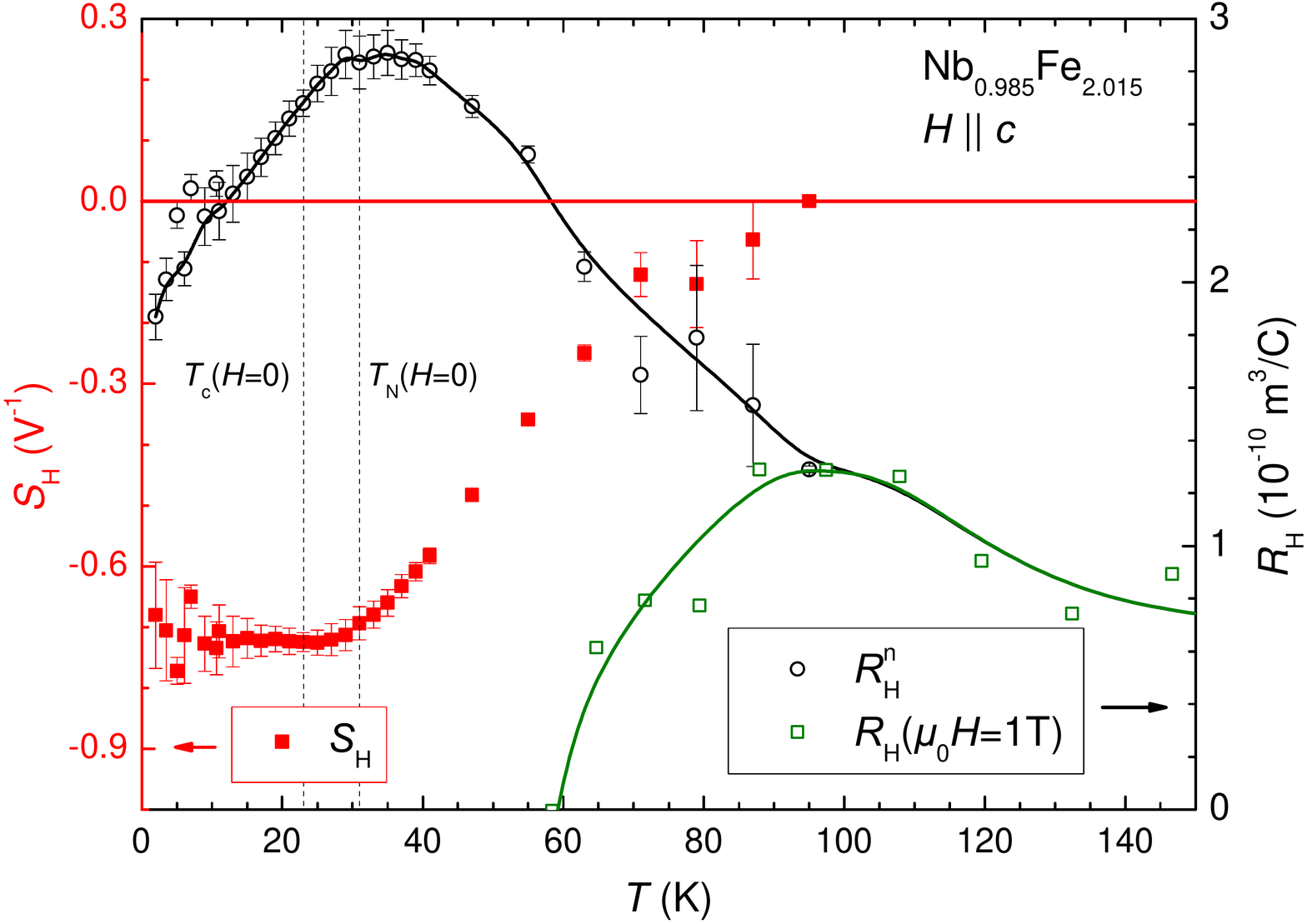}%
\caption{Normal and anomalous contribution as extracted from fits of Eq.~\ref{eq:AHEII} to $\rhoH(H)$ for \NbFeFe\ (cf.\ Fig.~\ref{fig:RhoH} (a)). Open symbols reproduce \RH\ from Fig.~\ref{fig:RHvsT_all} measured at \SI{1}{\tesla} as shown in Fig.~\ref{fig:RHvsT_all}. Lines are guides to the eye. Dashed vertical lines reflect the (zero-field) transition temperatures \TN\ and \TC.}%
\label{fig:NbFe2_AHEI}
\end{figure}

The temperature dependence of the extracted normal Hall coefficient and coefficient of the intrinsic AHE are displayed in Fig.~\ref{fig:NbFe2_AHEI}. Both \RHn\ and \SH\ obey a significant temperature dependence above \TN: \RHn\ increases from $\approx \HEunit{1}$ at \SI{150}{\kelvin} to almost \HEunit{3} at \SI{35}{\kelvin} and passes a maximum at \TN\ eventually decreasing to a value \HEunit{2} at lowest temperatures. Within a free-electron one-band model this would correspond to a charge carrier concentration of 2.5 electrons per formula unit at room temperature which decreases to slightly more than one electron per formula unit at lowest temperatures.

The comparison of \RHn\ with the as measured temperature dependence of $\RH(T)$ for all the samples depicted in Fig.~\ref{fig:RHvsT_all} indicates a common normal contribution throughout the \NbFey\ series: The observed maximum in $\RH(T)$ measured at a fixed field of \SI{1}{\tesla} is altered in its magnitude and position through the influence of the AHE which reduces the measured Hall coefficient at low temperature below the value obtained from \RHn\ alone.
For Fe-rich \NbFeFe\ the AHE is most pronounced and here the maximum value of $\RH(T)$ is most diminished. As the anomalous contribution becomes smaller for samples with less Fe content this maximum is more pronounced and extends to lower temperatures. This indicates that the electronic structure undergoes only minor changes for \NbFey.

 \SH\ is negligible at high temperatures, in accordance with the fact that $\rhoH(H)$ is linear in this regime. \SH\ becomes negative below \SI{100}{\kelvin} with a continuous increase of its absolute value. Only below \TN\ saturation sets in for \SH\ at a value of $\approx \SI{-0.7}{\per\volt}$ (cf.\ dashed line in Fig.~\ref{fig:NbFe2_AHEI}).

The constancy of \SH\ below \TC\ underpins the description in terms of the intrinsic AHE. Equivalent behaviour has been observed in the ferromagnetic phase of MnSi \cite{Lee2007,Neubauer2009a}. The value observed for \NbFeFe\ is very similar to that of MnSi in Ref.~\onlinecite{Neubauer2009a} ($\SH \sim \SI{-0.2}{\per\volt}$). The fact that the AHE in both \NbFeFe\ and MnSi is dominated by intrinsic contributions is reassuring of the regimes established for a variety of ferromagnetic systems, as both fall into the class of metallic systems with moderate impurity concentration \cite{Miyasato2007}.  For MnSi, however, the description failed at temperatures above \TC. Consequently, \NbFeFe\ displays a unique metallic system in which \SH\ continuously diminishes above \TC. The temperature dependence of the intrinsic AHE is to be contrasted to colossal magnetoresistance manganites for which \SH\ peaks  above \TC\ and obeys no constancy within the ferromagnetic phase \cite{Ye1999}.

For SDW-ordered \NbFe\ and \NbFeNb\ Eq.~\ref{eq:AHEII} is not able to reproduce the profile of $\rhoH(H)$. This is obvious from the fact, that in these samples the Hall resistivity obeys a minimum and a maximum. This sequence cannot be generated by two terms in Eq.~\ref{eq:AHEII} because both the resistivity and the magnetisation evolving monotonically in field. 
Qualitative agreement can be obtained by including a skew scattering contribution with opposite sign, but we have not been able to account quantitatively for the observed field dependence of the Hall resistivity in stoichiometric \NbFe\ and Nb-rich \NbFeNb\ (Fig.~\ref{fig:NbFe2_AHEI} (d) and (g)) over the entire temperature range.
\section{Conclusion}
The Hall effect in \NbFey\ comprises normal and anomalous contributions. For ferromagnetic \NbFeFe\ the anomalous contribution is covered by intrinsic AHE. This is in contrast to heavy-fermion systems of similar residual resistivities, for which skew-scattering dominates the AHE as shown for  Ce- and Yb-based compounds \cite{Fert1987,Paschen2005,Friedemann2010c}. In \NbFeFe, \rhoH\ can be modelled also above \TC\ where it provides a rare example for the complete suppression of the intrinsic AHE at elevated temperatures.  The intrinsic AHE may be compared to first principle band structure calculations \cite{Yao2004} and can be used to scrutinize different explanations of itinerant magnetism in \NbFe\ \cite{Tompsett2010,Subedi2010,Neal2011}. By contrast, for stoichiometric \NbFe\ and Nb-rich \NbFeNb\ an additional contribution to the AHE is identified. This additional contribution might be related to changes of the magnetic order and the associated magnetic fluctuations. 
Other possible explanations for such an additional contribution may consider the role of frustration in \NbFe, which features Kagom\'e layers of Fe atoms\cite{Taguchi2001}, or refined first principle calculations of the band structure as in \Molek{SrRuO}{3} \cite{Fang2003}.




%
%
\begin{acknowledgments}
The authors would like to thank S. Wirth and S. Paschen for fruitful discussions. SF acknowledges support by the Alexander von Humboldt foundation.
\end{acknowledgments}


\end{document}